\documentclass{article}
\usepackage[final]{nips_2017}
\usepackage[utf8]{inputenc} 
\usepackage[T1]{fontenc}    
\usepackage{hyperref}       
\usepackage{url}            
\usepackage{booktabs}       
\usepackage{amsfonts}       
\usepackage{nicefrac}       
\usepackage{microtype}      
\usepackage{graphicx}
\usepackage{enumitem}

\title{Applicability of Large Corporate Credit Models to Small Business Risk Assessment}

\author{
  Khalid El-Awady\thanks{} \\
  Department of Computer Science, Stanford University\\
  \texttt{kae@stanford.edu} \\
}

\begin{document}


\vspace{-0.15in}
\maketitle
\vspace{-0.3in}
\begin{abstract}
There is a massive underserved market for small business lending in the US with the Federal Reserve estimating over \$650B in unmet annual financing needs. Assessing the credit risk of a small business is key to making good decisions whether to lend and at what terms. Large corporations have a well-established credit assessment ecosystem, but small businesses suffer from limited publicly available data and few (if any) credit analysts who cover them closely. We explore the applicability of (DL-based) large corporate credit risk models to small business credit rating. 
\end{abstract}

\section{Introduction}	
A small business is one that has fewer than 1,500 employees and a maximum of \$38.5 million in average annual receipts, according to the Small Business Administration. Small businesses, though, often lack the size, assets (for collateral), financial history, or data that are typically used by traditional financial institutions to assess credit worthiness. While lenders extended nearly \$650B worth of loans to small businesses in 2019 \cite{SBA2019}, the Federal Reserve reports this represents less than half of these businesses credit needs \cite{KCFed2021}.

Large public businesses, on the other hand, provide a rich credit dataset. SEC rules require these businesses to publicly report detailed financial data in the form of profit and loss, balance sheet, and cash flow statements. Three main ratings agencies publish credit ratings for public companies: Standard \& Poor's (S\&P), Moody's, and Fitch.

We  study the applicability of deep learning-based credit models derived from large public corporate data to small businesses. We create a DL-based model to predict the credit ratings of large public companies from their financial statement data. We then explore the applicability of the model in forecasting adverse events or the probability of default of a small business on loan payments.

\section{Related work}
Traditionally, small business lenders rely on credit scores from one three credit reporting agencies: Dun \& Bradstreet, Experian, and Fico \cite{fundbox}. These utilize  business longevity, revenues, debt, owners' personal credit history, public records, and industry category, to come up with a risk score. 

Recently a number of new approaches have emerged. Divvy and Brex pioneered the use of real-time business bank balance monitoring to extend short term credit \cite{HBS2019}. Flowcast uses logistic regression and trees to estimate a risk score from business ERP transactional data such as invoices, shipments, and payment history \cite{Flowcast2018}. Visa released a new small business credit scoring service that uses logistic regression to estimate risk based on Visa payment transaction history \cite{Visa2020}. It is not known how well these proprietary models perform. They are also very invasive: businesses must consent to provide access to intimate business details for the score to be generated. 

In the realm of DL modeling of large public corporate credit risk the work of Golbayani et.al. \cite{GolbayaniMar2020} stands out as the most current and comprehensive study. The authors analyze DNN models on a dataset similar to ours. Their research suggests that DNNs perform well. They also explore variations on feature engineering and model training approaches that are useful. The work does not address applicability outside of the training dataset. We take this work as our starting point. Since the authors have not released their datset or code, we will build our dataset and models from scratch.

\section{Dataset and Features}
\subsection{The Labels}
\vspace{-0.1in}
\begin{figure*}[h]
    \centering
    \includegraphics[scale=0.4]{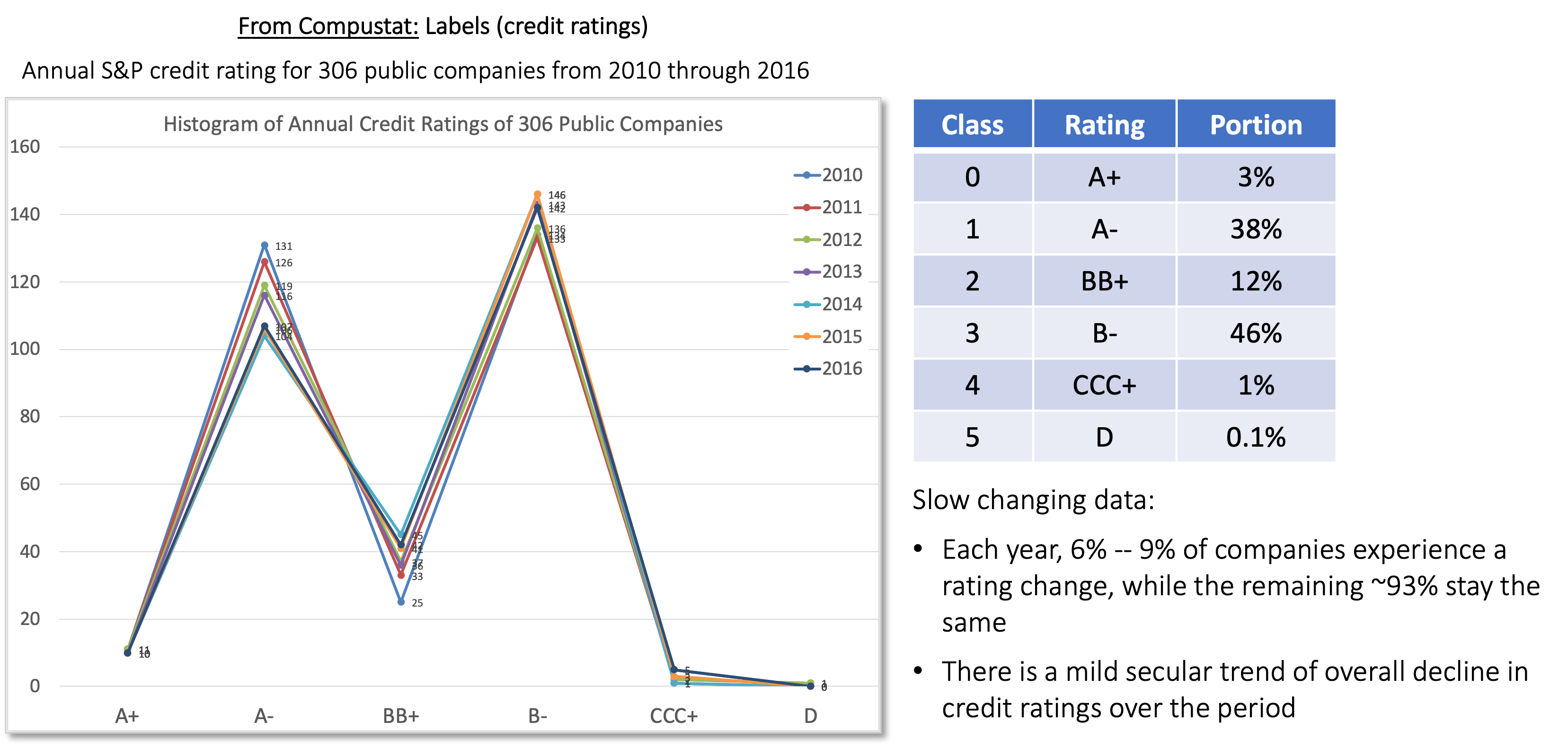}
    \caption{Histogram of annual long term credit ratings for 306 public companies from 2010 - 2016.}
    \label{CreditRating}
\end{figure*}

Our labels are S\&P corporate credit ratings, accessible from Compustat's database via Stanford Libraries \cite{Compustat}. We identified 306 public companies with annual long term credit ratings over the 7 year period period 2010-2016. 

Figure (\ref{CreditRating}) shows the credit ratings histogram. Ratings lie on a letter scale with '+' and '-' enhancements that includes up to 21 'grades', from 'AAA' (highest rating - extremely strong likelihood of repayment), down to 'D' (repayment default). Only 6 ratings (classes) actually appear in or data: 'A+', 'A-', 'BB+', 'B-', 'CCC+', and 'D'. The distribution is bi-modal and exhibits strong class imbalance. The rating change very slowly with only 7\% of companies on average experiencing a rating change in any given year. 

\subsection{The Inputs (Features)}
\vspace{-0.1in}
\begin{figure*}[h]
    \centering
    \includegraphics[scale=0.3]{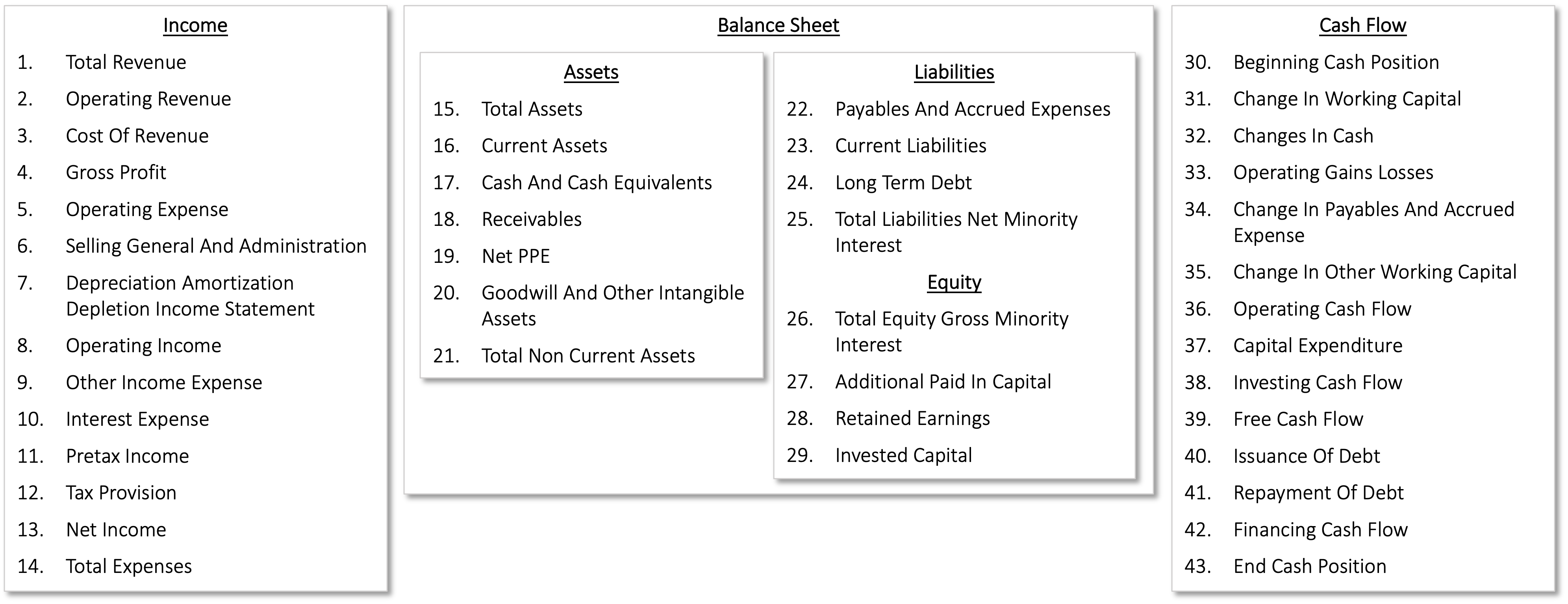}
    \caption{Input financial statement feature set for credit rating modeling.}
    \label{FeatureSet}
\end{figure*}

Access to Compustat's company financial data is restriced to Stanford GSB members only. Therefore we obtain  financials that form our features from Yahoo! Finance. There are $\sim$250 data fields available across all 3 financial statements, though each company uses only a subset. We analyzed our companies to determine a set of 43 of the most common fields that also comprise a coherent and complete set of financials, shown in figure (\ref{FeatureSet}). We excluded companies missing any of these fields over the time period. This resulted in 236 usable companies and a total of $m = 7 \times 236 = 1,652$ samples. Since financial data varies by orders of magnitude across companies (e.g., revenues from \$M to \$B) as well as across features, each feature is normalized to zero mean, unit variance across the sample set. 

\section{Methods}
\vspace{-0.1in}
\begin{figure*}[h]
    \centering
    \includegraphics[scale=0.35]{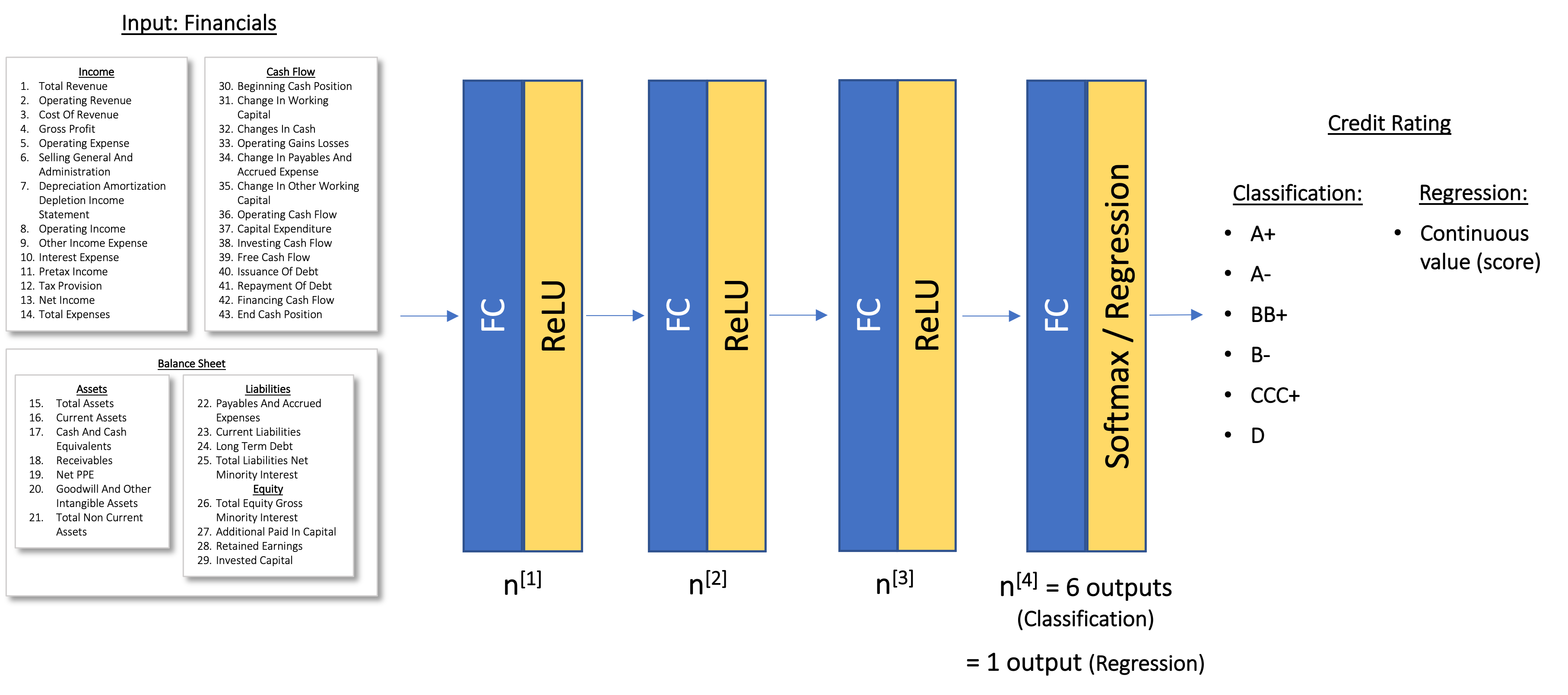}
    \caption{MLP model architecture with 3 hidden layers.}
    \label{MLP1HiddenLayer}
\end{figure*}
Our main model is two variations of a NN architecture. The first is classification-based and mimics the work of Golbayani et.al. \cite{GolbayaniMar2020}. Those authors thoroughly explore networks of different architectures (MLP, CNN, LSTM) and sizes, achieving accuracies of roughly 75\%-80\% on test and 80\%-85\% on training. They show small improvements with CNN and LSTM architectures over  MLP but not a lot. 

Our main goal is to test applicability of these models on a different dataset, so we seek a 'good enough' model. We show in the Experiments section that a 3 hidden-layer MLP using sparse categorical cross entropy loss does the job. This is similar to the main model of \cite{GolbayaniMar2020, Huang2004}. Adding batch normalization / dropout did not have a significant performance improvement and in some cases reduced accuracy, so it was omitted. Improvements with LSTM models were hard to obtain due to the mostly static ratings, though we postulate with more data better performance might be achievable. 

Since the credit classes lie on a continuum, we can modify this network to produce a risk score by replacing the softmax output activation by a single FC node. This regression-based approach is our second architecture. The classes are represented as sequential integers, and a mean squared error loss function is employed. This allows predictions to extend beyond the limits of the model classes, and the MSE loss also does a better job of penalizing deviations from nearby classes. 

For our relatively small sample, we randomly split the samples into 80\% for training (1,322 samples) and 20\% for testing (330 samples). To address the large class imbalances we employ the oversampling technique, SMOTE, to create a training set with equal number of samples in each class \cite{Chawla}.

\begin{figure*}[h]
    \centering
    \includegraphics[scale=0.42]{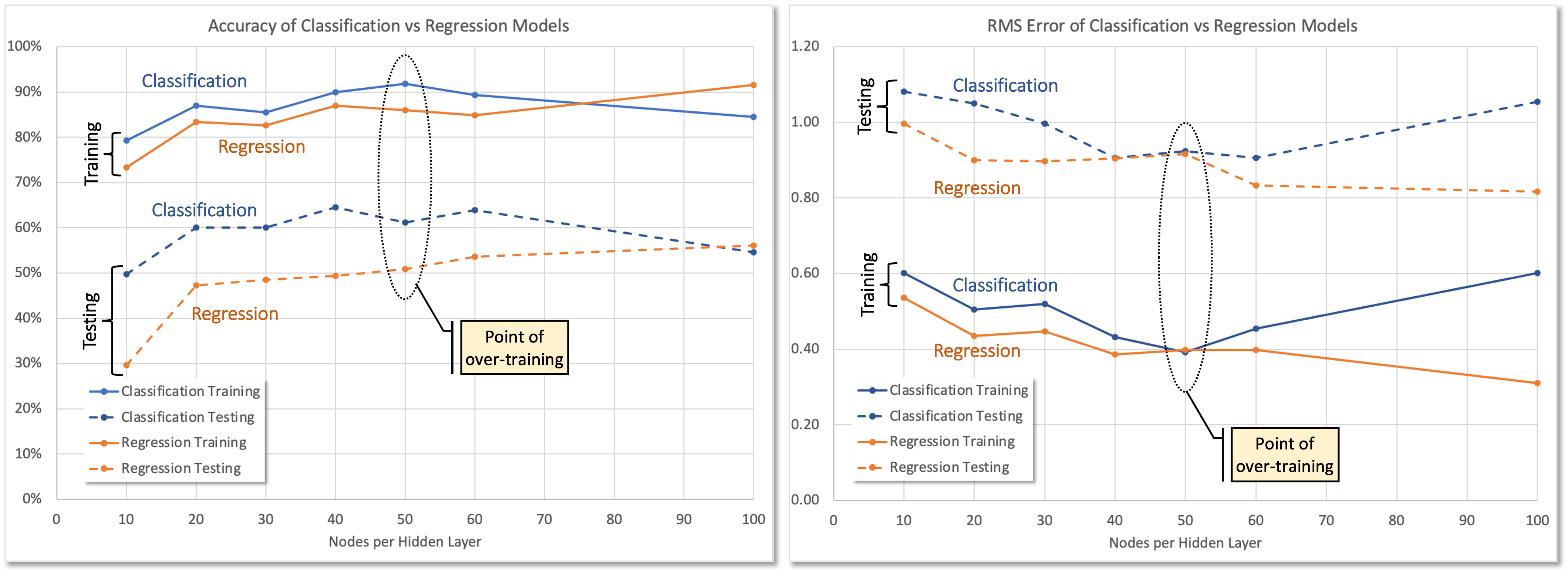}
    \caption{Accuracy \& RMS error for MLP with 3 hidden layers for varying number of nodes/layer.}
    \label{AccuracyRMS}
\end{figure*}

\section{Experiments/Results/Discussion}
\subsection{Multi Layer Perceptron (MLP) Model}
We assess the performance of our model with various nodes per layer as shown in figure (\ref{AccuracyRMS}). As expected the classification model outperforms on accuracy whereas the regression model outperforms on MSE. The differences are significant, though not overwhelming. Beyond 50 nodes per layer, the performance of the classification model degrades, indicating over-training. We select 50 nodes per layer as our working model size. These results are obtained after 3,000 epochs of training with no batching, which experimentation showed worked well. 

\begin{figure*}[h]
    \centering
    \includegraphics[scale=0.37]{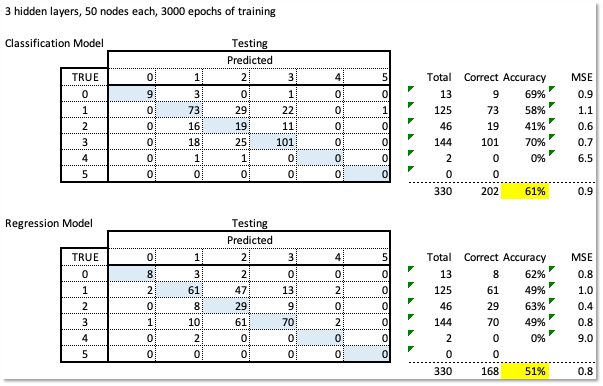}
    \caption{Comparison of the confusion matrices for classification and regression models.}
    \label{ConfusionMatrix}
\end{figure*}

The benefits of the regression model are further seen from the confusion matrix in figure (\ref{ConfusionMatrix}). Note how the regression model keeps mis-classifications closer to the main diagonal. 
\vspace{-0.1in}

\subsection{Testing Outside the Dataset: Bankruptcies}
\begin{figure*}[h]
    \centering
    \includegraphics[scale=0.39]{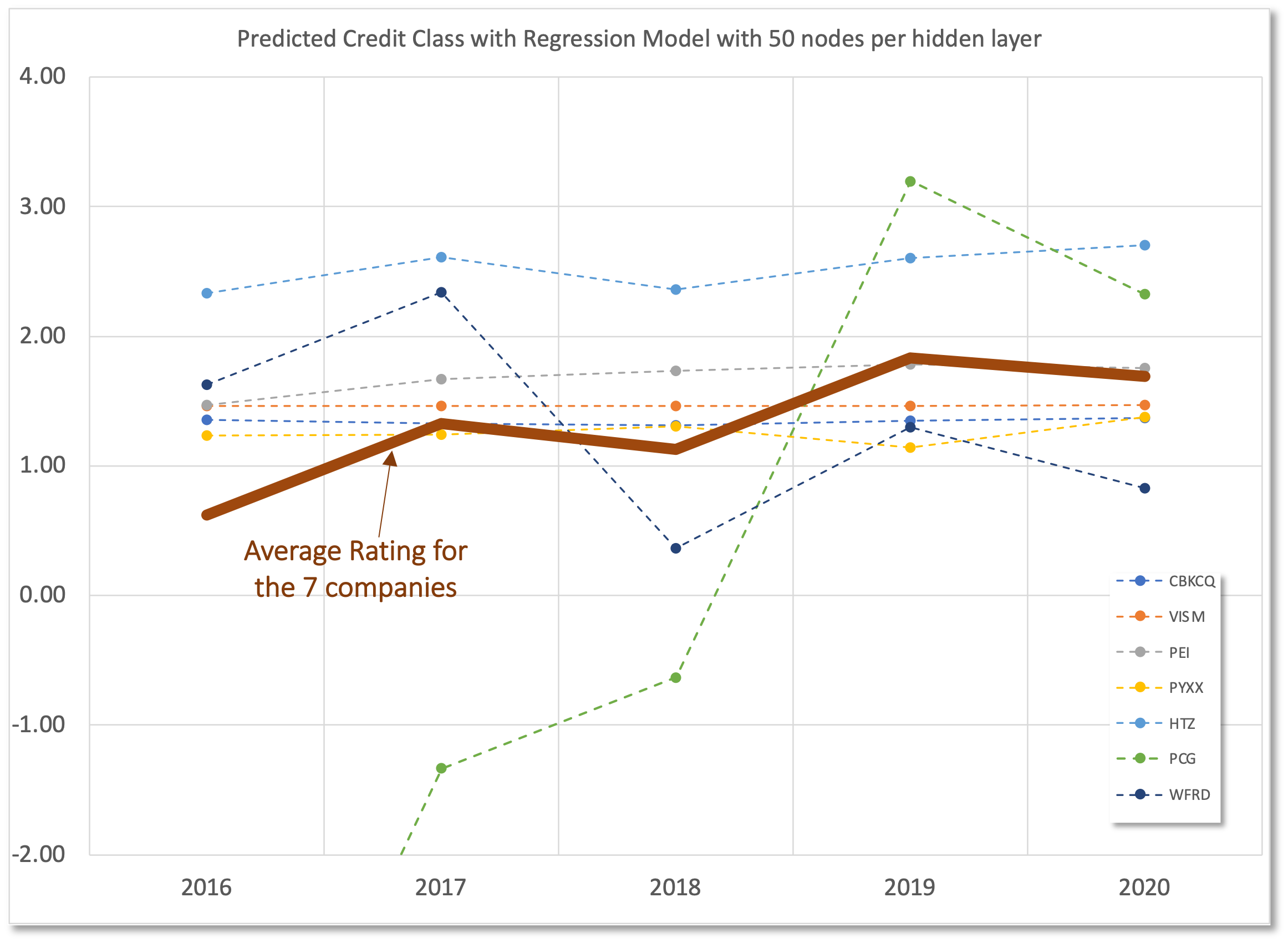}
    \caption{Predicted classes for companies that experienced a bankruptcy in 2019/2020 time frame.}
    \label{BankruptcyPrediction}
\end{figure*}
Our first model application outside the dataset is to test detection of adverse credit events. We apply the model to a set of public companies that filed Chapter 11 bankruptcy in 2019 or 2020. Using the list from S\&P and Wikipedia \cite{SPBankruptcy2020, WikipediaBankruptcy2019}, we extract 7 such companies that had a full feature set over the period 2016-2020. If our model is relevant, we expect to see a souring of credit as these companies approach the bankruptcy event (i.e., an increase in predicted class output). The results are shown in figure (\ref{BankruptcyPrediction}). Some of the companies show considerable rating volatility over the period and the average does exhibit an overall upward trend as would be expected. 

\subsection{Testing Outside the Dataset: Small Cap Businesses}
\begin{figure*}[h]
    \centering
    \includegraphics[scale=0.42]{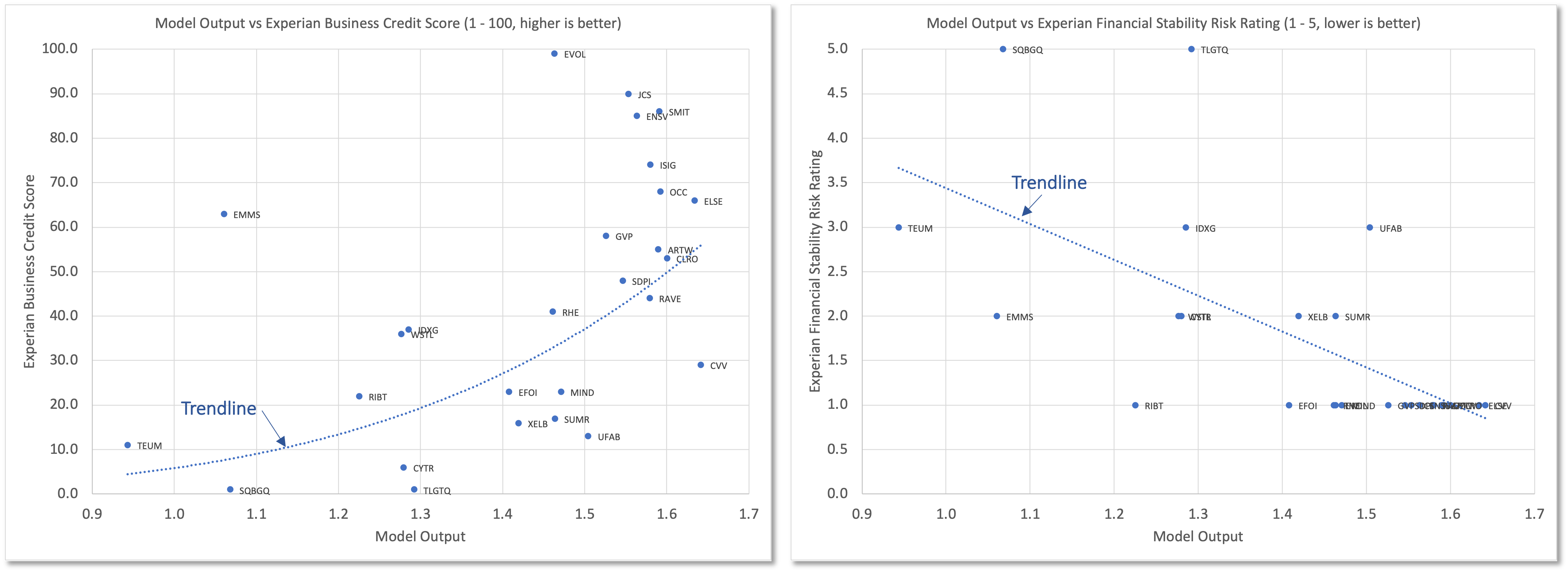}
    \caption{Comparsion of model output with Experian risk scores.}
    \label{ExperianComparison}
\end{figure*}

We use small cap public companies as a proxy for small businesses. Starting with the 100 smallest companies (by market cap) of the Wilshire 5,000 index, all of which have market caps under \$35M \cite{WilshireTail}, we find 27 companies with a full feature set for the year 2020. We compare the regression model output for these against the current value of 2 risk scores produced by Experian \cite{ExperianWebsite}. One score is their 'Business Credit Score' which ranges on a scale of 1-100, with higher score indicating lower risk, and it predicts the likelihood of a serious credit delinquency within the next 12 months. The other score is their 'Financial Stability Risk Rating' which ranges on a scale of 1-5, with a lower score indicating lower risk, and it predicts the likelihood of payment default within the next 12 months. We plot the output of our model against these scores in figure (\ref{ExperianComparison}).

The result is surprising. We would expect our model to be negatively correlated with the Business Credit Score and positively correlated with the Financial Stability Risk Rating, but we see the opposite. Assuming Experian's scores are meaningful then either our model is wrong, or the financial factors that make large corporate businesses more credit worthy have the opposite effect on small businesses. One other observation across all our tests is that the model generates scores in a very narrow output range. This is likely a result of the input normalization using the mean and standard deviations derived from the training set on large companies.  

\section{Conclusion/Future Work }
Credit risk modeling is murky. While a number of agencies are recognized as rating authorities, the underlying models they use are proprietary and opaque. Further, these models have known deficiencies \cite{Partnoy2017}. Yet these models are widely used and represent the best available measures of corporate credit risk. DNN's can approximate the predictions of these models relatively well, though not entirely, using just financial data. And while these models show some predictive capability outside the dataset, it is not clear how significant these are. 

Many extensions of this work are possible. I would like to undertake a model interpretability exercise. Classical finance relies heavily on certain financial statement ratios in assessing credit risk \cite{Ganguin2004}. Our model may confirm some of these are key predictors, or unearth new ones that perform better. Our regression model also needs refinement with more sophisticated architectures and an expansion of the datasets (which were tedious to collect and could be expanded with better access to the right databases.) Finally, with enough data from Experian (we had a limit of 30 companies) we could train models to fit those scores directly, or combine that data with our dataset to train a single model. 

\section{Contributions}
\begin{itemize}[leftmargin=*]

    \item Presented a classification NN that approximates the credit risk assessment of large public companies reasonably well, confirming that financials play a large part in driving credit risk, but not everything. 
    
    \item Extended the model to a regression one that outputs a continuous score instead of just a rating class. 
    
    \item Investigated the use of our regression score model to detect adverse events outside the original dataset. We showed that there is correlation between or model output and bankruptcy events. 
    
    \item Investigated the use of our regression score model against a proprietary small business risk score from Experian. We show negative correlation between the two models. This warrants further investigation whether our model needs improvement or the financial factors that are beneficial to large corporate credit ratings are somehow detrimental to small businesses.
    
\end{itemize}

\bibliographystyle{plain}

\end{document}